\documentclass[showpacs,preprintnumbers,amsmath,amssymb,twocolumn,aps,pre]{revtex4}

\usepackage{epsfig,graphics,graphicx,amssymb}

\begin{document}

\title{The role of demographic stochasticity in a speciation model with sexual reproduction}
\author{Luis F. Lafuerza and Alan J. McKane}

\affiliation{Theoretical Physics Division, School of Physics and Astronomy, The University of Manchester, Manchester M13 9PL, UK}

\date{\today}

\begin{abstract}
Recent theoretical studies have shown that demographic stochasticity can greatly increase the tendency of asexually reproducing phenotypically diverse organisms to spontaneously evolve into localised clusters, suggesting a simple mechanism for sympatric speciation. Here we study the role of demographic stochasticity in a model of competing organisms subject to assortative mating. We find that in models with sexual reproduction, noise can also lead to the formation of phenotypic clusters in parameter ranges where deterministic models would lead to a homogeneous distribution. In some cases, noise can have a sizeable effect, rendering the deterministic modelling insufficient to understand the phenotypic distribution.
\end{abstract}
\pacs{05.40.-a, 87.23.Cc, 87.10.Ca} 

\maketitle

\section{Introduction}
Establishing the determinants of biological diversity is a fundamental question in biology. A particular aspect of this question that has attracted a great deal of attention is the distribution (in genotype or phenotype space) of a population of interacting organisms, and to what extent and under what conditions clusters of similar individuals tend to arise~\cite{GavriletsBook,MaynardSmithBook,Doebeli210}. A better understanding of this question could shed some light onto the process of sympatric speciation, whereby a `mother' species splits into two or more other species without geographic isolation \cite{Coyne07,Bolnick07}.

There is a history of mathematical models proposed to elucidate the mechanisms whereby species (as clusters of phenotypically similar organisms) tend to form spontaneously due to competition~\cite{McarthurLevins67,Sasaki97,Noest97,DieckmannDoebli99,krenke03}. Models initially considered asexual populations, to facilitate analytical tractability. It was shown that spontaneous clustering can occur~\cite{Sasaki97,DieckmannDoebli99,Scheffer06}, but the phenomena is somewhat sensitive to particular assumptions about the functional form of the interaction kernels used~\cite{Pigolotti07,Pigolotti10}, questioning the biological relevance of the findings. Models in which reproduction is sexual tend to lead to spontaneous clustering with less restrictive assumptions~\cite{Noest97,Doebli07}, but the biological realism of the conditions required has also been questioned~\cite{Bolnick07}. Recently, it was shown that stochastic effects can greatly increase the range of parameters for which species are formed in asexual models~\cite{RogersMckanepl,RogersMckanebiol}, presenting stochastic pattern formation as a novel mechanism for speciation, which has particular biological implications~\cite{1mmspecies}. In this paper we will study the effects of stochasticity in a speciation model with sexual reproduction. We will show that demographic stochasticity can increase the parameter range in which species clustering is observed and that, in some cases, noise can have a sizeable effect, rendering the deterministic modelling insufficient to understand the phenotype distribution.

The rest of the paper is organised as follows. We will present the model as well as its mathematical formulation in Sec.~\ref{sec:Model}. In Sec.~\ref{sectiondet} we will present the analysis in the deterministic (large population) limit, summarising some previously known results and presenting some new ones. We will then perform the analysis of the stochastic effects in Sec.~\ref{sec:Stochastic}, highlighting the cases in which noise effects are largest. We will conclude with a summary and conclusions. Some technical details are left for two appendixes.

\section{Description of the model \label{sec:Model}}
The model consists of a population of individuals which reproduce sexually and die through competition with each other. Each individual is described by its phenotype, which determines the extent to which the organism competes with other organisms, and the likelihood with which it can mate with a given other organism. For simplicity, we will assume that the phenotype is well described by a single scalar variable which can take on all possible real values. A simple example in which a single scalar variable is the main determinant of the strength of competition between organisms could be the beak size in birds or the jaw size in lizards~\cite{RoughgardenBook} (which determines the extent to which they feed on the same resources), or in general, body size. The model is stochastic in that deaths and births are random events (which, however, take place with probabilities determined by the state of the system). 
There are two basic processes:
\begin{itemize}
\item[(i)] \textit{Death}.\,\,The death rate (probability per unit time) of individual $i$, $d_i$, is given by $d_i= [Kf(x_i)]^{-1}\,\sum_jg(x_i-x_j)$, where $g(x)$ is the competition kernel that quantifies the strength with which two individuals with phenotype distance $x$ compete, $K$ is a constant that controls the overall carrying capacity of the ecosystem and $f(x)$ is a function that determines the relative intrinsic advantage of phenotype $x$.
\item[(ii)] \textit{Reproduction}. Each individual reproduces at a rate one. The probability that individual $i$ mates with individual $j$ is proportional to $m(x_i-x_j)$, with $m$ a function determining the strength of assortment (preference for individuals which are alike). If individuals $i$ and $j$ mate, an offspring is generated with phenotype given by $x_0=(x_i+x_j)/2+\zeta$, where $\zeta$ is a random variable with probability density function $r(x)$. This models mutation about the average of the parents phenotype $(x_i+x_j)/2$.
\end{itemize}

This simple reproduction rule is justified when the character under consideration is determined by a large number of additive genes (that is, without dominance or epistasis~\cite{BulmerBook}); in this limit, the `reproduction noise', $\zeta$, is a Gaussian random variable with zero mean and variance given by $\sigma^2_m$, that, for simplicity, will be assumed to be phenotype-independent (the influence of this quantity will be one of the main aspects of our investigation). The function $f(x)$ modulates the relative advantage of the phenotypes; it typically decays to zero for large $|x|$, constraining the phenotypes to a given region of interest.
We are also assuming that the organisms are hermaphroditic, but this assumption will be relaxed later. The model is a generalisation of the asexual model studied in~\cite{RogersMckanepl}. 

Following~\cite{RogersMckanepl}, we will describe the state of the system by the density in phenotype space, $K\phi(x)$, where
\begin{equation}
\phi(x)=\frac{1}{K}\sum_{i=1}^{N(t)}\delta(x-x_i).
\label{phi_density}
\end{equation}
Here $N(t)$ is the number of organisms at time $t$ and $\delta(x)$ is the Dirac delta function. We have introduced the carrying capacity, $K$, to obtain a function $\phi(x)$ that has a well-defined $K\rightarrow\infty$ limit.

When an organism with phenotype $y$ dies, $\phi(x)$ is modified by the subtraction of delta function centred at $y$. Similarly, if a new organism with phenotype $y$ is born, $\phi(x)$ is modified by the addition of a delta function at $y$. With this in mind, we define the operators $\Delta_y^{\pm}$ by their action on a generic functional $F[\phi(x)]$ as:
\begin{equation}
\Delta_y^{\pm}F[\phi(x)]=F[\phi(x)\pm \frac{1}{K}\delta(x-y)].
\end{equation}
Now suppose that $\gamma(x,\phi)$ is the density rate at which an individual with phenotype $x$ dies, so that $\gamma(x,\phi)dxdt$ is the probability that an individual with phenotype in the interval $(x,x+dx)$ dies in the time interval $(t,t+dt)$, given that the state of the system is given by $\phi$ at time $t$. The definition of the process implies that $\gamma(\phi,t)$ is given by:
\begin{eqnarray}
 \gamma(x,\phi)&=&K\phi(x)\frac{1}{Kf(x)}\sum_{i=1}^{N(t)}g(x-x_i)\nonumber\\
&=&K\frac{\phi(x)}{f(x)}\int\phi(y)g(x-y)dy \nonumber \\
&\equiv& K\frac{\phi(x)}{f(x)} \phi * g(x).
\label{gamma_def}
\end{eqnarray}
 
The probability that individual $i$, if it mates, does so with individual $j$ is $m(x_i-x_j)/\sum_km(x_i-x_k)=m(x_i-x_j)/K\int\phi(y)m(x_i-y) dy = m(x_i-x_j)/K \phi * m(x_i)$. The probability density that their offspring has phenotype $x$ is given by $r(x-(x_i+x_j)/2)$, where $r(x)$ is the Gaussian probability density with variance $\sigma^2_m$. With this in mind, we see that the rate density at which a new individual is created with phenotype $x$, $\beta(x,\phi)$, is:
\begin{eqnarray}
 &&\beta(x,\phi)=\sum_{i,j=1}^{N(t)} r(x-(x_i+x_j)/2)\frac{m(x_i-x_j)}{K \phi * m(x_i)}\label{birthrate}\\
&=&K\int \phi(y)\int\frac{\phi(z)m(y-z)}{\phi * m(y)}r(x-(y+z)/2)dydz. \nonumber
\end{eqnarray}
Combining the two contributions to the change in $\phi$, the probability density of finding the system at state $\phi$ at time $t$, $P(\phi,t)$, changes in time according to the following functional master equation \cite{RogersMckanepl}:
\begin{eqnarray}
 \frac{\partial}{\partial t}P(\phi,t)&=&\int[(\Delta_x^{-}-1)\beta(\phi,x)P(\phi,t)\nonumber\\
&+&(\Delta_x^{+}-1)\gamma(\phi,x)P(\phi,t)]dx.\label{ME}
\end{eqnarray}

Due to the non-linearity of the system (that arises due to the interactions) we are unable to obtain an exact solution and some approximation scheme is needed to proceed. Expanding the $\Delta^\pm_x$ operators in Eq.~(\ref{ME}) to second order in $K^{-1}$, we can derive a functional Fokker-Planck equation, given in Appendix~\ref{appFPE}. For our purposes, it is clearer to work with the equivalent stochastic differential equation which takes the form (see Eqs.~(\ref{SDE_general}) and (\ref{corr_gen}))
\begin{eqnarray}
& & \frac{\partial \phi(x,t)}{\partial t} = -\int\phi(x,t)\phi(y,t)\frac{g(x-y)}{f(x)}\,dy \nonumber \\
&+& \int\frac{\phi(y,t)\phi(z,t)}{\phi * m(y)}m(y-z)r(x-(y+z)/2)\,dy\,dz \nonumber \\
& & + \frac{\eta(x,t)}{\sqrt{K}},
\label{SDE_specific}
\end{eqnarray}
where $\eta(x,t)$ is a Gaussian white noise with zero mean and with a correlator which is given by Eq.~(\ref{corr_gen}) of the Appendix. It is interesting to compare this equation with the analogous equation in the asexual case. There $m(x-y)=\delta(x-y)$ and so $\phi * m(x) = \phi(x,t)$. Then Eq.~(\ref{SDE_specific}) becomes
\begin{eqnarray}
& & \frac{\partial \phi(x,t)}{\partial t} = -\int\phi(x,t)\phi(y,t)\frac{g(x-y)}{f(x)}\,dy \nonumber \\
&+& \int \phi(y,t) r(x-y)\,dy + \frac{\eta(x,t)}{\sqrt{K}},
\label{SDE_asexual}
\end{eqnarray}
which is the equation found in Ref.~\cite{RogersMckanepl}, apart from the function $f(x)$, which was not included in the form of the model previously analysed (note that we are allowing self-fertilization since, for simplicity, in Eq. (\ref{birthrate}) we do not exclude $i=j$; forbidding the $i=j$ case would add $O(1/K)$ terms). We will now analyse Eq.~(\ref{SDE_specific}), first of all in the deterministic limit, and then in the general stochastic setting.

\section{Deterministic analysis}\label{sectiondet}
The deterministic limit corresponds to taking $K\rightarrow\infty$, and so the governing equation is simply Eq.~(\ref{SDE_specific}), but with the last (noise) term absent. Some progress may be made analytically if we assume that the ecological functions, namely the competition kernel [$g(x)$], the mating function [$m(x)$], the function modulating the carrying capacity [$f(x)$] and the offspring distribution [$r(x)$], are all Gaussian functions. We denote their variances by $\sigma^2_c, \sigma^2_a, \sigma^2_f$ and $\sigma^2_m$, respectively. In this case, the deterministic equation has a Gaussian stationary solution, $\phi(x)_{\rm st}=Ce^{-x^2/(2\sigma^2_{\rm st})}$, with $\sigma^2_{\rm st}$ and $C$ both satisfying complicated algebraic equations; the equation for $\sigma^2_{\rm st}$ being first derived by Doebeli et.~al.~\cite{Doebli07}. These authors also found that numerical integration of the deterministic equation showed that when $\sigma^2_c$ and $\sigma^2_m$ are small compared with $\sigma^2_f$, there is an intermediate range of $\sigma^2_a$, $\sigma^2_m\lesssim\sigma^2_a\lesssim\sigma^2_f/3-\sigma^2_m$, for which the Gaussian solution becomes unstable and a multi-modal stationary solution is obtained. 

The random mating case ($\sigma^2_a\rightarrow\infty$) is particularly interesting to analyse. In this limit, the equation for $\sigma^2_{\rm st}$ reduces to a cubic equation $\sigma^6_{\rm st} + \alpha_4 \sigma^4_{\rm st} + \alpha_2 \sigma^2_{\rm st} + \alpha_0 = 0$, where
\begin{eqnarray}
& & \alpha_4 = 2\sigma^2_m + \sigma^2_c, \ \ \alpha_2 = \sigma^2_f\sigma^2_c - 2\sigma^2_m(2\sigma^2_f-\sigma^2_c), \nonumber \\
& & \alpha_0 = - 2\sigma^2_m\sigma^2_c\sigma^2_f.
\label{alpha_defns}
\end{eqnarray}
Since $\alpha_4 > 0$ and $\alpha_0 < 0$, the cubic equation always has a single positive solution. This then is the required solution. In this limit, the constant $C$ is found to be
\begin{equation}
C=\sqrt{2\frac{\sigma^2_{\rm st}+\sigma^2_c}{\sigma^2_{\rm st}+2\sigma^2_m}},
\label{C_equation}
\end{equation}
(we have not taken $f(x)$ to be normalised; $f(x)=\exp(-x^2/(2\sigma^2_f))$, so that $\sigma^2_f$ controls the size of phenotype space available). If we now, in addition, investigate the $\sigma^2_f\rightarrow\infty$ limit (the most relevant when the phenotype distribution is not too constrained by the external fitness landscape), we find two very different regimes, depending on the relative width of the competition kernel and the reproduction noise distribution. When the reproduction noise, $\sigma^2_m$, is smaller than a critical value given by $\sigma^2_c/4$, the variance of the phenotype distribution is small and independent of $\sigma^2_f$, $\sigma^2_{\rm st}=2\sigma^2_m\sigma^2_c/(\sigma^2_c-4\sigma^2_m)$. If $\sigma^2_{\rm st}$ depends on $\sigma^2_f$ in a way which means that it diverges as $\sigma^2_f\rightarrow\infty$, then the term $\sigma^6_{\rm st}$ dominates over the term quartic in $\sigma_{\rm st}$ and the term quadratic in $\sigma_{\rm st}$ dominates over $\alpha_0$. Therefore, $\sigma^2_{\rm st}\simeq\sqrt{\sigma^2_f(4\sigma^2_m-\sigma^2_c)}$, and this implies that if the reproduction noise exceeds the critical value, $\sigma^2_m>\sigma^2_c/4$ then the variance of the phenotype distribution grows linearly with $\sigma_f$. In both cases, numerical integration of the deterministic equation shows that the Gaussian distribution is always stable and the phenotype distribution is always uni-modal, a consequence of the random mating.

The factors determining the transition into a multi-modal distribution can be understood in a simpler way if we assume that the range of phenotype space is finite, for instance $-\pi < x \leq \pi$, and assume that the deterministic equation (Eq.~(\ref{SDE_specific}) without the noise term) satisfies periodic boundary conditions. The competition function, $g$, the assortment function, $m$, and the offspring distribution will be assumed to be periodic functions of period $2\pi$. Since there is now no need for the function $f(x)$ to regulate the competition process at large $|x|$, we set $f(x)=1$. While the periodic boundary condition assumption is biologically unrealistic, we expect it to have a small impact
when the scales of the competition, mating and offspring distributions are all much smaller than the available region of phenotype space (as determined by $f$).

Under these conditions, the deterministic equation has a uniform solution $\phi(x,t) =$ constant. Using the normalisation $\int^{\pi}_{-\pi} g(x)\,dx = 1$, one finds that $\phi_{\rm st} = 1$. A linear stability analysis of the deterministic equation around the solution $\phi = 1$ is performed in Appendix~\ref{appFPE}. This shows that the uniform solution is unstable if:
\begin{equation}
 2 r_km_{-k/2} - r_km_km_{-k/2} - 1 - g_k>0,
\label{linearstab}
\end{equation}
for some $k$, with $r_k, m_k$ and $g_k$ the Fourier modes of the reproduction noise, the mating and the competition kernel, respectively. An equivalent expression to (\ref{linearstab}) in a slightly different model was derived in~\cite{Noest97}. 

\begin{figure}[h]
 \includegraphics[scale=0.4]{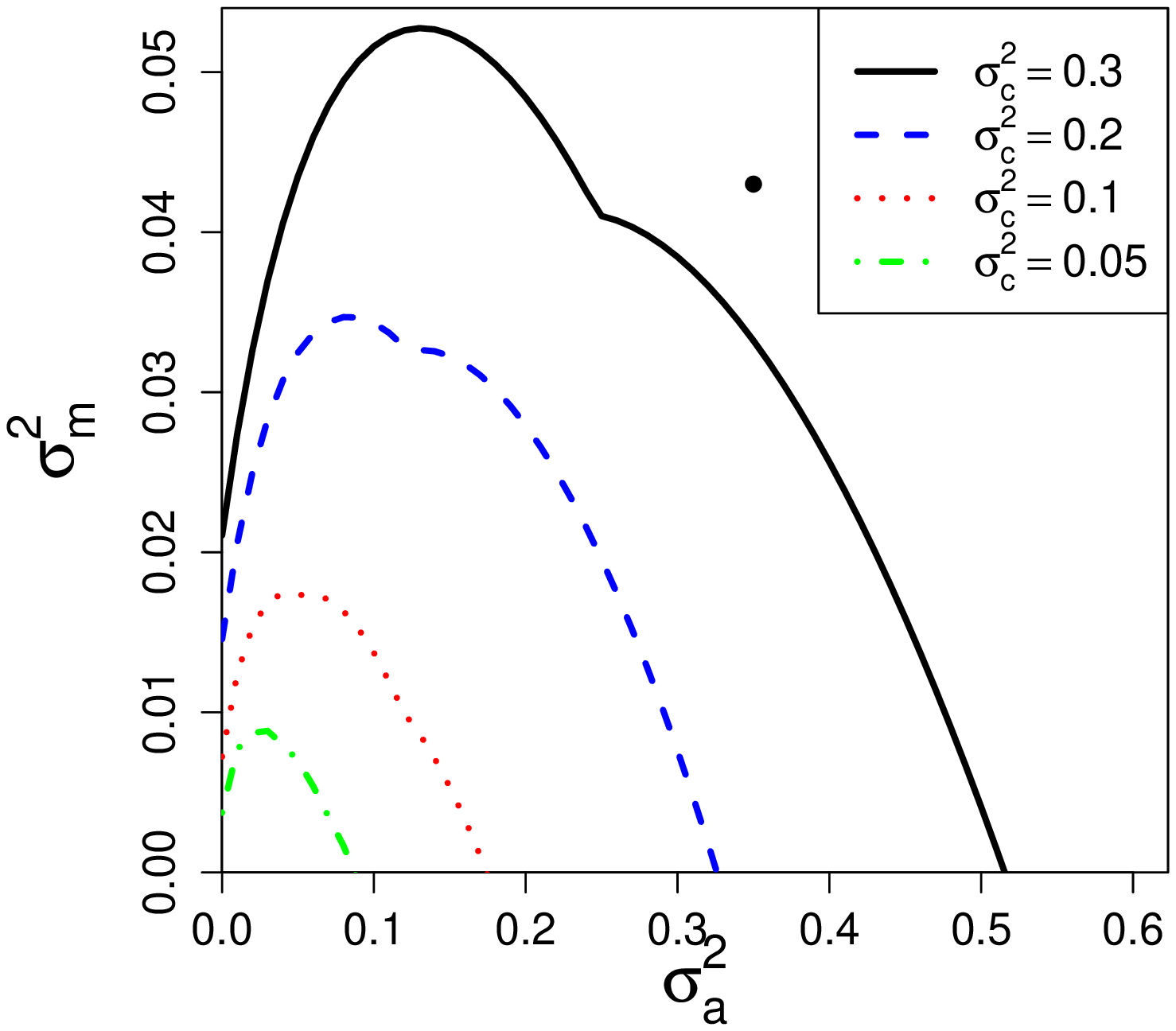}
 \includegraphics[scale=0.4]{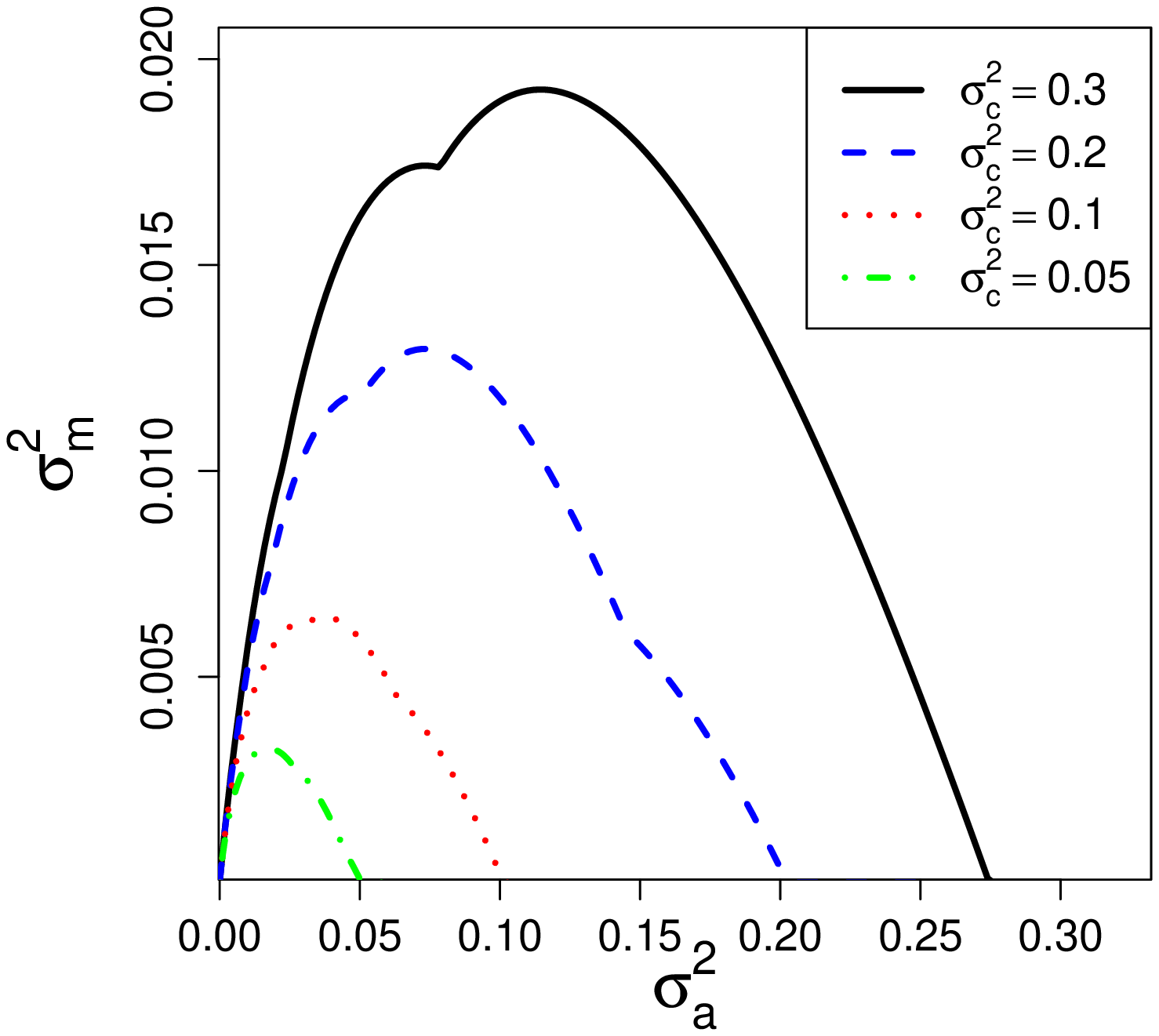}
\caption{Region of stability of the homogeneous solution of the deterministic equation in $\sigma^2_a-\sigma^2_m$ space (mating range - mutation range), for several values of $\sigma^2_c$ (competition range), for uniform (upper panel) and Gaussian (lower panel) forms of the ecological functions. The homogeneous solution is unstable below the lines, leading to the appearance of patterns. The kinks are produced when the maximum value of the left-hand side of the inequality (\ref{linearstab}) changes from one value of $k$ to another. The black dot marks parameter values used in Fig.~\ref{figstochclus}.}\label{bifdia}
\end{figure}

From the inequality (\ref{linearstab}) one can obtain the bifurcation diagram of the system that shows the regions of parameter space in which the deterministic equation [Eq.~(\ref{detfield})] leads to patterns. Figure~\ref{bifdia} portrays several bifurcation diagrams for the case in which the ecological functions are uniform and Gaussian functions (projected onto the $(\sigma^2_a,\sigma^2_m)$ plane). The figure shows that phenotypic clusters are observed for low values of the reproduction noise $\sigma_m$ and for intermediate values of the assortativity scale $\sigma_a$, in accordance with what was found in \cite{Doebli07}. Intuitively, clusters tend to form with a distance, $d$, between them so that they barely compete (so $d \gtrapprox\sigma_c$), but these clusters will only be durable if individuals from different clusters do not mate (so $d>\sigma_a$), suggesting that clusters will not form if $\sigma_a\gg\sigma_c$. Sexual reproduction tends to concentrate the phenotype distribution, so a moderate value of $\sigma_a$ can promote the formation of clusters. Moreover, the clusters can be well-defined only if the reproduction noise is not too large ($\sigma_m<d$). 


In the asexual case~\cite{RogersMckanepl,RogersMckanebiol} it was found that the range of parameters for which patterns occur was much greater than that predicted by a stability analysis of the kind carried out above. So motivated by the expectation that the same may be true in the case of sexual reproduction, we go on to analyse our model for finite values of $K$, when stochastic fluctuations will be present.

\section{Stochastic effects}\label{sec:Stochastic}
\subsection{Weak noise effects}\label{weak}
\begin{figure}[h]
  \includegraphics[scale=0.47,angle=0]{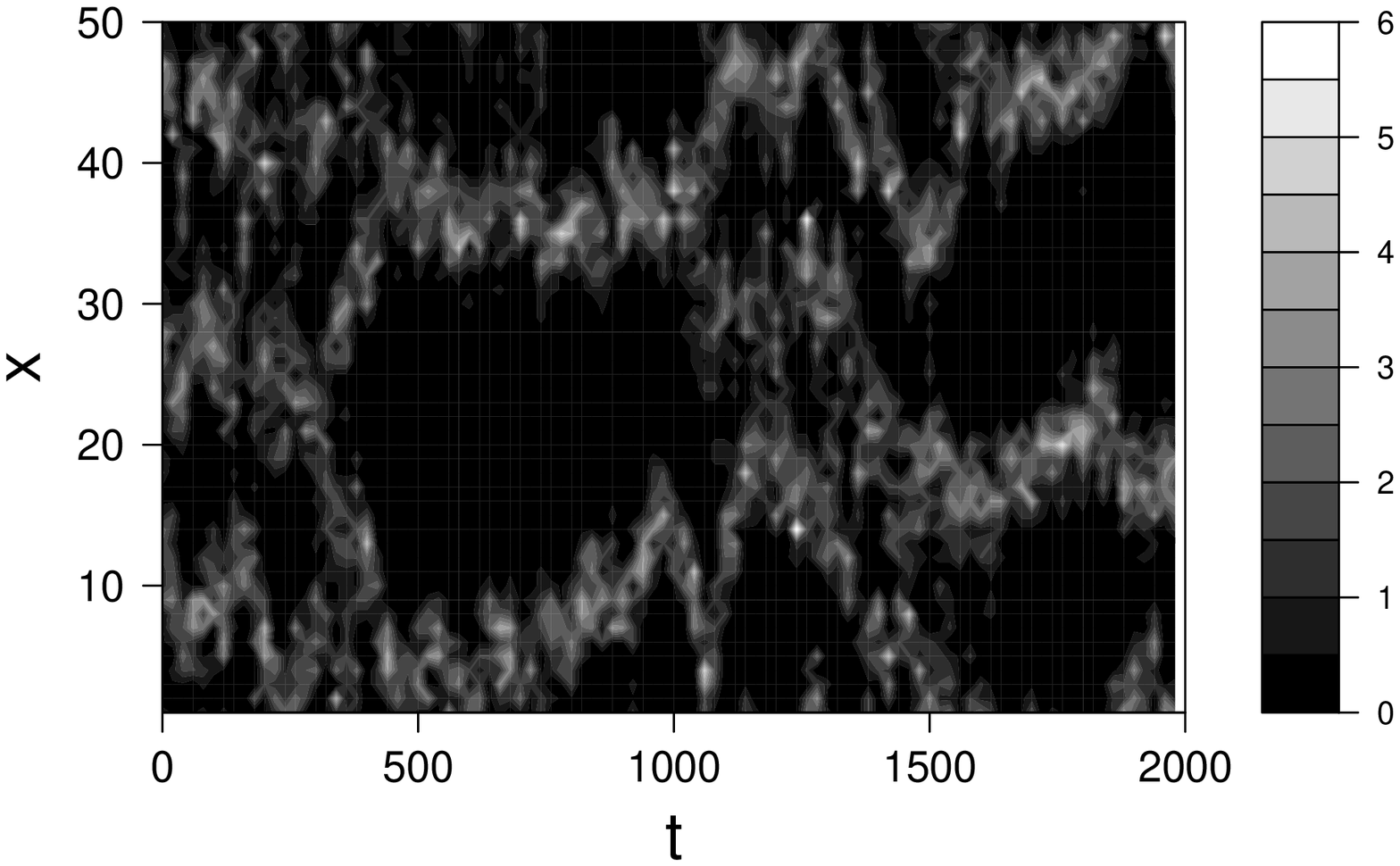}
  \includegraphics[scale=0.47,angle=0]{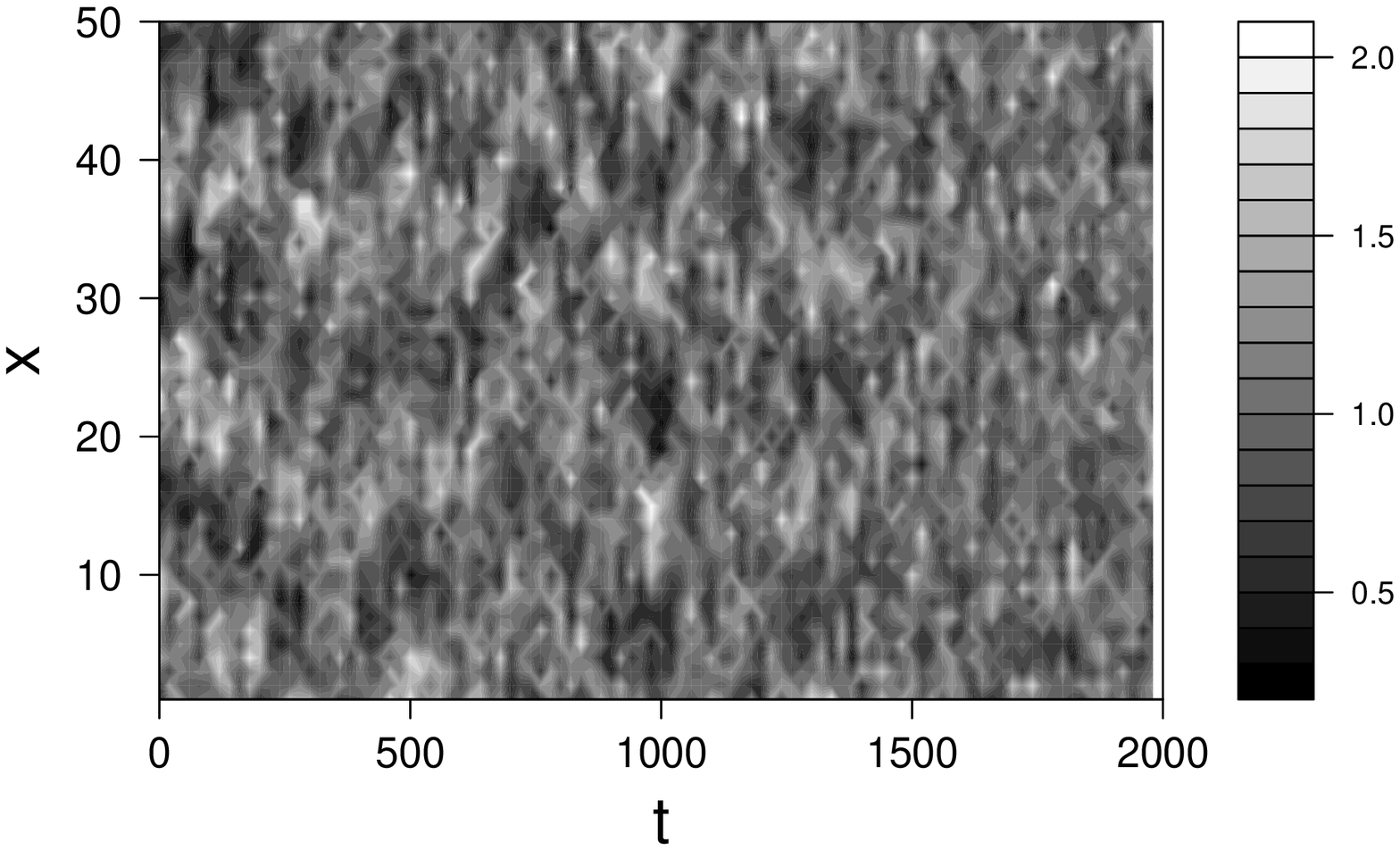}
\caption{Time evolution of the phenotype distribution for $\sigma^2_c=0.3, \sigma^2_{a}=0.35, \sigma^2_{m}=0.043$ and uniform ecological functions, in the region of deterministic stability of the homogeneous solution (marked as a black dot in Fig.~\ref{bifdia}). The carrying capacity is $K=20$ (upper panel) and $K=200$ (lower panel). Note that for the smaller carrying capacity clear phenotypic clusters form.}\label{figstochclus}
 \end{figure}

Numerical simulations of the individual-based model show that, for moderate values of the carrying capacity, $K$, the observed phenotype distribution does not always agree with the results obtained from the deterministic equation (\ref{detfield}). For example, at points in parameter space corresponding to a stable uniform solution, but not too far from the instability boundary, one can observe clusters in phenotype space forming for small values of $K$ (see Fig.~\ref{figstochclus}).


The origin of these stochastic patterns can be understood by analysing the stochastic differential equation (\ref{SDE_specific}). Specifically, we apply the van Kampen system-size expansion~\cite{vankampen2007} by expanding $\phi$ about the homogeneous solution found in Sec.~\ref{sectiondet} and writing $\phi(x,t)=1+K^{-1/2}\xi(x,t)$. The factor $K^{-1/2}$ reflects the nature of the fluctuations at large $K$, and $\xi(x,t)$ is a new stochastic field. The bulk of the calculation is exactly the same as that carried out when performing the linear stability analysis in the deterministic case, except in this case the existence of the $K^{-1/2}$ factor ensures that the noise term in Eq.~(\ref{SDE_specific}) is retained. Thus, going over to Fourier variables and using Eq.~(\ref{LSA_result}), one finds that
\begin{eqnarray}
\frac{d }{dt} \xi_k(t)&=&\left[2 r_k m_{-k/2} -1 - g_k -r_k m_k m_{-k/2} \right] \xi_k(t)\nonumber\\
&+& \eta_k(t),
\label{LNA_result}
\end{eqnarray}
where $\eta_k(t)$ is the Fourier transform of $\eta(x,t)$. The correlation function of this noise is only calculated to leading order within the linear noise approximation, that is, setting $\phi(x,t)=1$. This gives, using Eqs.~(\ref{A_and_B}), (\ref{Psi_and_Xi}) and (\ref{corr_gen}), $\int \mathcal{B}(\phi, x, y)\,dy = 2$, and therefore
\begin{equation}
\langle \eta_k(t) \eta_{-k}(t') \rangle = 2\left( 2\pi \right)\delta(t - t'),
\label{corr_k}
\end{equation}
showing that, as usual, the noise is additive within the linear noise approximation.

\begin{figure}[h]
  \includegraphics[scale=0.48]{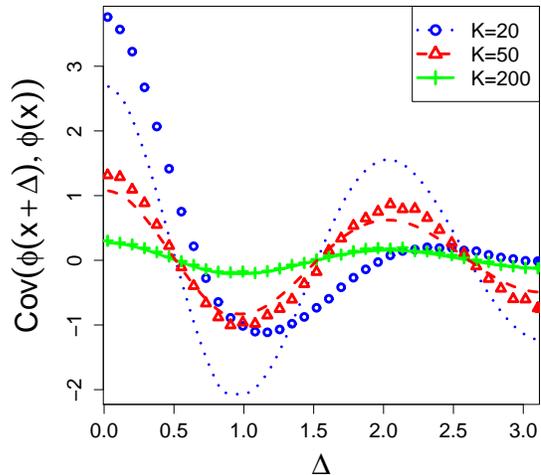}
 \caption{Spatial covariance of the phenotype density distribution. The ecological functions are uniform distributions with variances given by $\sigma^2_c=0.3$, $\sigma^2_a=0.35$, $\sigma^2_m=0.043$ (marked as a black dot in Fig. \ref{bifdia}). Numerical results (symbols) were averaged over at least 100 measurements, after a transient of $t=2000$. The lines correspond to the theoretical expression, Eq.~(\ref{covariance}).}\label{figcov}
 \end{figure}

The size of the stochastic patterns can be quantified by looking at the spatial covariance of the phenotype distribution in the stationary state:
\begin{eqnarray}
& & \text{Cov}(\phi(x),\phi(x+\Delta)) \equiv \nonumber \\
& & \left\langle\left(\phi(x)-\langle\phi(x)\rangle\right)\left(\phi(x+\Delta)-\langle\phi(x+\Delta)\rangle\right)\right\rangle.
\end{eqnarray}
If the distribution in phenotype space shows high-density regions separated by low-density ones with a well-defined average distance, the spatial covariance will display a sizable spatial modulation in $\Delta$.
In the linear noise regime, the covariance becomes 
\begin{eqnarray}
&&\int\text{Cov}(\phi(x),\phi(x+\Delta))dx=\frac{1}{K}\int\langle\xi(x)\xi(x+\Delta)\rangle dx \nonumber \\
&&=\frac{1}{2\pi K}\sum_{n}\langle\xi_n\xi_{-n}\rangle e^{in\Delta}.
\end{eqnarray}
Using Eq.~(\ref{stationary_state_result}), and assuming that $g_n, m_n$ and $r_n$ are all even functions of $n$ (which follows if the ecological functions are symmetric), we may find $\langle\xi_n\xi_{-n}\rangle$, and then finally obtain:
\begin{eqnarray}
& & \int\text{Cov}(\phi(x),\phi(x+\Delta))dx \nonumber \\
& & =\frac{1}{K}\sum_{n=-\infty}^{\infty}\frac{e^{in\Delta}}{\left[ 1+g_n+r_nm_nm_{n/2}-2r_nm_{n/2} \right]}.
\label{covariance}
\end{eqnarray}
Expression (\ref{covariance}) is compared against numerical simulations in Fig.~\ref{figcov} for the case in which the ecological functions are uniform distributions. For low values of the carrying capacity, $K$, the covariance shows a clear spatial modulation, revealing the presence of stochastic patterns. The theoretical expression agrees qualitatively, and the quantitative agreement becomes better as $K$ increases, where the linear noise assumption is expected to be a better approximation to the stochastic dynamics. The results are very similar when the ecological functions have other forms, for instance if they are Gaussians or belong to the family $\exp(-|x|^l/(2\sigma^l))$ with varying $l$.


In summary, the linear noise approximation shows that stochastic pattern formation can lead to the formation of clusters in phenotype space in situations when the deterministic description predicts a uniform distribution. Interestingly, the stochastic patterns (as opposed to the deterministic ones) decrease with the carrying capacity (that controls the abundance of individuals), which leads to particular biological implications \cite{1mmspecies} when this mechanism is dominant.

When comparing with the asexual case~\cite{RogersMckanepl}, the stochastic patterns now appear on a relatively narrow zone close to the deterministic transition line. We will now show that stochastic effects can also play a more prominent role in the sexual case.

\subsection{Strong noise effects}\label{strong}
\begin{figure}[h]
\includegraphics[scale=0.4,angle=0]{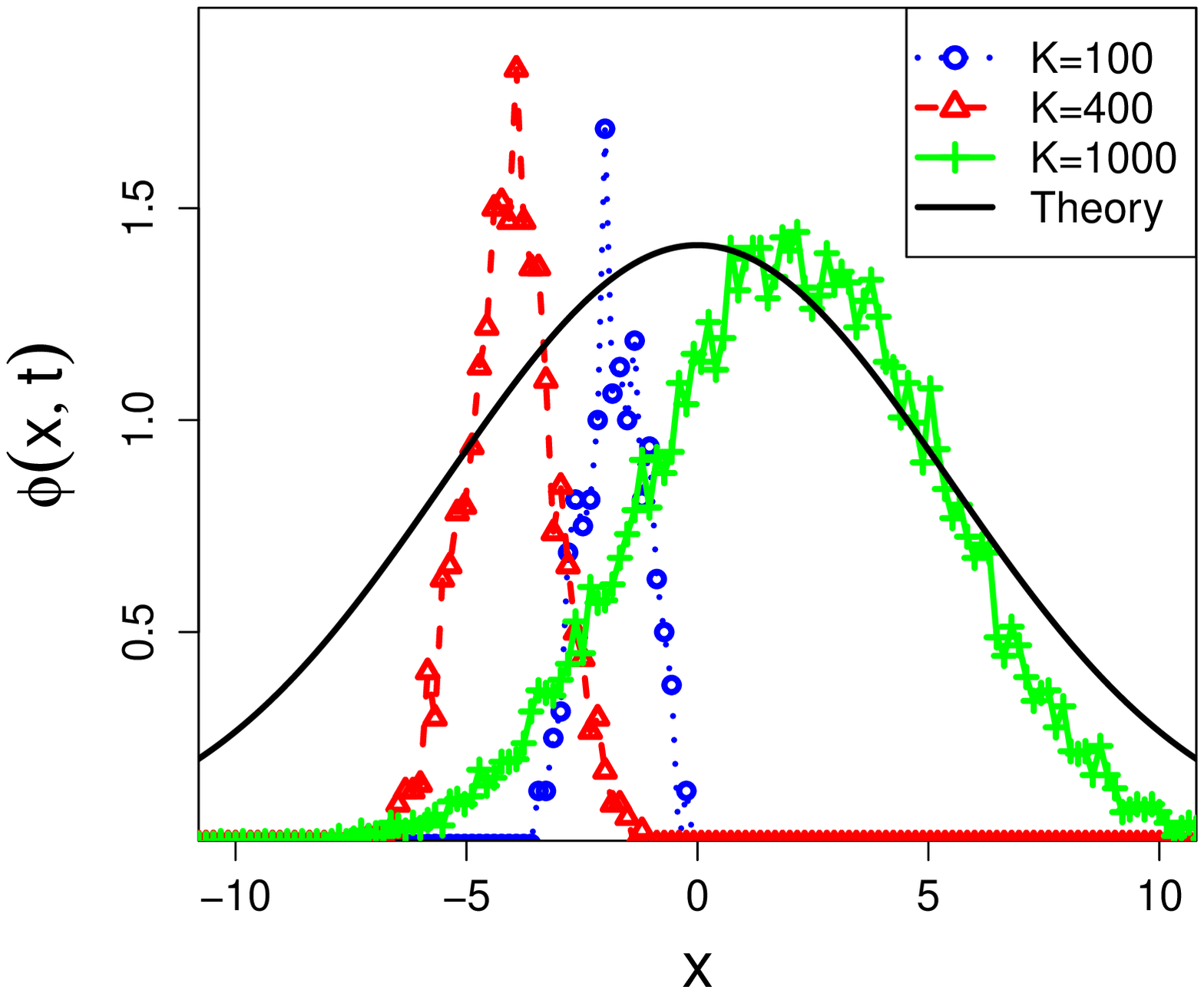}
\includegraphics[scale=0.4,angle=0]{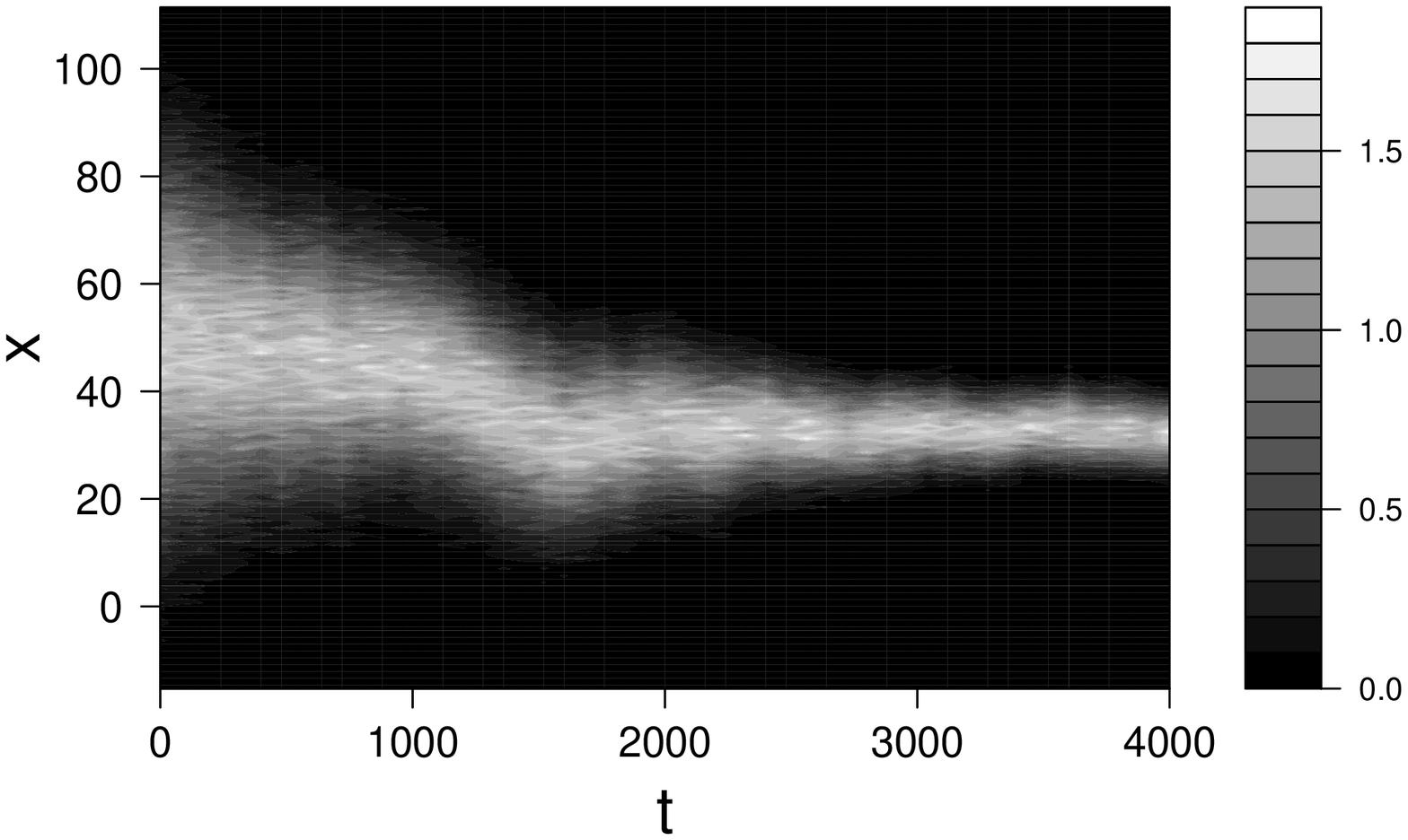}
\caption{Upper panel shows the observed instantaneous phenotype distribution in the steady state for different values of the carrying capacity, together with the deterministic prediction. Lower panel shows the evolution of the phenotype distribution for $K=400$, starting with the stationary distribution predicted by the deterministic analysis. Parameter values are $\sigma^2_{c}=0.02, \sigma^2_{m}=0.04,\sigma^2_{f}=6400$.}\label{figgaussians}
\end{figure}

A stronger noise-induced effect occurs when the mating is random. As discussed in Sec.~\ref{sectiondet}, in the random mating case there are two regimes: one leading to a narrow phenotype distribution, for low reproduction noise ($4\sigma^2_{m}<\sigma^2_c$), and one leading to a broad phenotype distribution in the large reproduction noise case. In the broad phenotype regime ($4\sigma^2_{m}>\sigma^2_c$), simulations show that the phenotype distribution observed is much narrower that the one predicted by the deterministic analysis, see Fig.~\ref{figgaussians}. This effect happens rather generically in this regime.


In the random mating case, since the mating range is the largest scale of the system, assuming periodic boundary conditions gives results that are quite different from those of the system with open boundaries. In this last case there is a strong bias towards the centre of the phenotypic space. For this reason we focus on the open boundary conditions case, which is the more biologically relevant. This, however, greatly complicates the mathematical analysis of the noise effects. Simulations show (Fig. \ref{figgaussians}) that stochastic effects lead to the formation of a phenotypic cluster with a width that approaches the deterministic prediction only for very large values of the carrying capacity, $K$. We, therefore, find that in this regime demographic stochasticity has a quite large effect, again leading to the formation of tight phenotypic clusters when the deterministic analysis predicts a broad distribution. Similar results are obtained when the ecological functions do not have a Gaussian form, showing the robustness of this phenomenon.

\section{The case when individuals belong to one of two sexes}
\label{two_sexes}
So far we have assumed, for simplicity, that any two organisms can mate, i.e. the organisms are hermaphroditic. We now go on to model the situation with two explicitly different sexes. 

We will denote the phenotype of female organism by $x_i$ and that of the male organism by $y_\alpha$. We use different indices, since the number of male and female organisms at a given time will typically be different, and so the range of these indices will be different. The model is as in Sec.~\ref{sec:Model}, but with the following modifications:
\begin{itemize}
\item[(a)] Only females reproduce and at rate one.
\item[(b)] The probability that female organism $i$ mates with male organism $\alpha$ is proportional to $m(x_i-y_\alpha)$. 
\item[(c)] The probability that the offspring is male or female is $1/2$.
\item[(d)] The competition is independent of sex, so that the death rate of organism $i$ is $d_i=\sum_j g(x_i-x_j)+\sum_\alpha g(x_i-y_\alpha)$.
\end{itemize}

With these assumptions, one can follow through the analysis given in Sec.~\ref{sectiondet} and find the analogues of Eq.~(\ref{SDE_specific}):
\begin{eqnarray}
& & \frac{\partial \phi_f}{\partial t} = - \int\phi_f(x,t)\left[\phi_f(y,t)+\phi_m(y,t)\right]\,\frac{g(x-y)}{f(x)}\,dy \nonumber \\
&+& \frac{1}{2} \int\frac{\phi_f(y,t)\phi_m(z,t)}{\phi_m * m(y)}m(y-z)r(x-(y+z)/2)\,dy\,dz \nonumber \\
& & + \frac{\eta_f(x,t)}{\sqrt{K}},
\label{SDE_female}
\end{eqnarray}
and
\begin{eqnarray}
& & \frac{\partial \phi_m}{\partial t} = - \int\phi_m(x,t)\left[\phi_f(y,t)+\phi_m(y,t)\right]\,\frac{g(x-y)}{f(x)}\,dy \nonumber \\
&+& \frac{1}{2} \int\frac{\phi_f(y,t)\phi_m(z,t)}{\phi_m * m(y)}m(y-z)r(x-(y+z)/2)\,dy\,dz \nonumber \\
& & + \frac{\eta_m(x,t)}{\sqrt{K}},
\label{SDE_male}
\end{eqnarray}
where the subscripts $f$ and $m$ denote female and male respectively. It is convenient to work with the sum and differences of $\phi_f$ and $\phi_m$: $S(x,t)=\phi_f(x,t) + \phi_m(x,t)$ and $D(x,t)=\phi_f(x,t) - \phi_m(x,t)$. If we additionally take $K \to \infty$ to obtain the deterministic equations, we find that
\begin{eqnarray}
& & \frac{\partial S(x,t)}{\partial t} = - \int S(x,t) S(y,t)\,\frac{g(x-y)}{f(x)}\,dy \nonumber \\
&+&\frac{1}{2} \int\frac{[S(y,t)+D(y,t)][S(z,t)-D(z,t)]}{[S-D] * m(y)}\nonumber \\
&&  \times \left[ m(y-z)r(x-(y+z)/2)\,dy\,dz \right],
\nonumber \\
\label{det_S}
\end{eqnarray}
and
\begin{equation}
\frac{\partial D(x,t)}{\partial t} = - \int D(x,t) S(y,t) \frac{g(x-y)}{f(x)}\,dy.
\label{det_D}
\end{equation}
From Eq.~(\ref{det_D}) we see that a steady state solution is $D(x)=0$, that is, $\phi_f(x) = \phi_m(x)$. Then the deterministic equations for $\phi_f$ and $\phi_m$ collapse into each other, and agree with the deterministic equation in the hermaphroditic case, apart from the factor of $1/2$ seen in Eqs.~(\ref{SDE_female}) and (\ref{SDE_male}). If, as in Sec.~\ref{sectiondet}, we take the ecological functions to be all Gaussians, we again find a Gaussian stationary solution $\phi(x)_{\rm st}=Ce^{-x^2/(2\sigma^2_{\rm st})}$, with $\sigma^2_{\rm st}$ and $C$ satisfying the same complicated algebraic equations, except that $C$ takes on a value of one quarter the value found in Sec.~\ref{sectiondet}. Note that the total population is now half that obtained in the hermaphroditic case because now we assume that only females can initiate reproduction events.

To examine the instability leading to the appearance of patterns, we can again assume a finite interval $-\pi < x \leq \pi$ for phenotypic space, with periodic boundary conditions. We can now look for homogeneous stationary solutions, and study the stability of these solutions. It is clear that once again $D=0$ is a stationary solution, with $\phi_f = \phi_m = 1/4$ under the same conditions discussed in Sec.~\ref{sectiondet}. Linearising about these homogeneous solutions, we write $S(x,t)=\frac{1}{2}+\tilde{S}(x,t)$ and $D(x,t)=\tilde{D}(x,t)$. Keeping only linear terms in $\tilde{S}$ and $\tilde{D}$, we have from Eq.~(\ref{det_D}) that
\begin{equation}
\frac{\partial \tilde{D}(x,t)}{\partial t} = - \frac{1}{2} \tilde{D}(x,t) \ \ \
\Rightarrow \ \tilde{D}(x,t) = \tilde{D}(x,0)\,e^{-t/2},
\label{tildeD}
\end{equation}
which shows that the solution $D(x,t)=0$ is always stable. For $\tilde{S}(x,t)$, it is more convenient to work in Fourier space (see Appendix~\ref{appFPE}). We find that
\begin{eqnarray}
\frac{d\tilde{S}_k(t)}{dt} &=& \frac{1}{2} \left[ 2 r_km_{-k/2} - r_km_km_{-k/2} - 1 - g_k \right] \tilde{S}_k(t) \nonumber \\
&+& F_k \tilde{D}_k (t),
\label{tildeS}
\end{eqnarray}
where $F_k$ is a function of $k$ only. Equation (\ref{tildeD}) shows that $\tilde{D}_k(t)$ decays exponentially with $t$, so Eq.~(\ref{tildeS}) implies that one obtains the same stability condition as the one found for $\phi$ in the hermaphroditic case, Eq.~(\ref{linearstab}). We can therefore conclude that the stability boundaries in the case of two sexes are identical to those found in the hermaphroditic case. This is confirmed by numerical simulations of the stochastic version of the model. We also find that stochastic pattern formation takes place as in the hermaphroditic case, with results from the case of two sexes being equivalent to those of the hermaphroditic case, but with a factor of $1/4$ in the carrying capacity.

\section{Conclusion}
\label{conclude}
The precise definition of exactly what constitutes a species is still open to debate~\cite{Mallet95,Coyne&Orr04}. One of several different alternatives is the `phenotypic clustering species concept', in which species correspond to distinct phenotypic clusters, analogous to Mallet's `genotypic clustering species concept'~\cite{Mallet95}. In this view, the clustering of individuals in trait or gene space that we recognise as species is a pattern that emerges from underlying ecological and evolutionary mechanisms. Just as mixtures of chemical constituents which react and diffuse may create patterns (for instance, spots and stripes), so individuals which react (for example, compete) and diffuse (for example, mutate in trait or gene space) may create patterns (clusters). The traditional approach of the theoretical physicist would then be to construct a simple model to see if the effect appears, and if so, then see if a deeper understanding of the effect can be gained from an analysis of the model.

This is the approach that we have adopted here. We have started from a simple model that only contained birth, death (through competition) and mutation, and asked under which conditions clusters of individuals in phenotype space were formed. As discussed in the Introduction, this is a question which has been investigated by several authors, however our study focused on individuals who gave birth only after mating with another individual, whereas most previous investigations assumed asexual reproduction.

We also investigated stochastic pattern formation~\cite{Black2012,McKane2014}. Most previous work was carried out in the case of infinite carrying capacity (in our notation, $K \to \infty$) where the governing equations are deterministic. The standard way of proceeding in this case is to look under what conditions the constant density (homogeneous) solution of this equation is unstable. This is carried out by performing a linear stability analysis about this homogeneous solution. However it has been found that frequently patterns are still found in regions of parameter space where the homogeneous solution of the deterministic equation is stable. These patterns typically are stochastic, and can be found by analysing the governing equations at finite carrying capacity, $K$. Interestingly, the scaling of these patterns with $K$ has particular biological implications \cite{1mmspecies}.

The search for stochastic patterns in models of asexual reproduction has been carried out previously~\cite{RogersMckanepl,RogersMckanebiol}. It was found that noise originating in the discrete nature of individuals can lead to the spontaneous formation of species in situations where this would not happen deterministically. The main purpose of this paper was to extend this to the case of sexual reproduction. We first supposed that individuals played both the male and female role (i.e. were hermaphrodite). We found that when mating is assortative (i.e. organisms show preference for like individuals) stochastic patterns can appear in the region of stability of the deterministic homogeneous solution. These patterns were, however, somewhat restricted to parameter values not too far from the deterministic instability boundary. When mating is random the stochastic effects are stronger, and the phenotype distribution for moderate $K$ is always relatively narrow, in contrast with the deterministic predictions. The case where the individuals were either male or female led to very similar results, which in some cases could be mapped directly onto the the hermaphroditic case.

There are several extensions of the current work which could be carried out. One could distinguish between ``genetic noise'' (arising from recombination and mutations), which affects the inheritable traits, and ``environmental'' or ``developmental'' noise, which leads to two individuals with same genotype to have different phenotypes and which is not inherited. These two types of noise are likely to have rather different effects in the phenotype distribution. Also, if a dominant part of the genetic noise is due to recombination, that is, the trait considered is determined by the effect of many genes in a diploid organism, and the effect of the different genes is additive (no epistasis or dominance), then the noise should depend on the position in genotype space (i.e. it would be multiplicative noise). Multiplicative noise would break the spatial symmetry and could lead to interesting effects, but would be more difficult to study analytically. Another possible extension is to include the Allee effect, which we have here ignored for simplicity. We believe, however, that the work discussed here shows that the inclusion of stochastic effects is vital if we wish to predict the range of parameters for which patterns, and therefore possibly species, occur.

\section*{Acknowledgments}
We wish to thank Tim Rogers for useful discussions. This work was supported in part by the EPSRC (UK) under grant number EP/H02171X.

\begin{appendix}

\section{The mesoscopic evolution equation}
\label{appFPE}

In the limit of large carrying capacity, the master equation (\ref{ME}) can be expanded in powers of $K^{-1}$ to give the following functional Fokker-Planck equation (analogous to the derivation in Ref.~\cite{RogersMckanepl} for the asexual case):
\begin{eqnarray}
 \frac{\partial}{\partial t}P(\phi,t)&=&-\int\int \frac{\delta}{\delta\phi(x)}\left[\mathcal{A}(\phi,x,y)P(\phi)\right]dxdy\label{FP}\\
&+&\frac{1}{2K}\int\int\frac{\delta^2}{\delta\phi(x)^2}\left[\mathcal{B}(\phi,x,y)P(\phi)\right]dxdy,\nonumber
\end{eqnarray}
where terms of $K^{-2}$ and higher in the expansion have been neglected. Here
\begin{eqnarray}
\mathcal{A}(\phi,x,y)&=& \Psi(\phi,x,y) - \Xi(\phi,x,y),\nonumber\\
\mathcal{B}(\phi,x,y)&=& \Psi(\phi,x,y) + \Xi(\phi,x,y),
\label{A_and_B}
\end{eqnarray}
where 
\begin{eqnarray}
\Psi(x,y) &=& \int\frac{\phi(y)\phi(z)}{\phi * m(y)}m(y-z)r(x-(y+z)/2)\,dz,\nonumber\\
\Xi(x,y)&=& \phi(x)\phi(y)\frac{g(x-y)}{f(x)}.
\label{Psi_and_Xi}
\end{eqnarray}

A completely equivalent way of expressing the stochastic dynamics of the system, is to write down the equivalent stochastic differential equation. This takes the form~\cite{gardiner2009,risken1989} 
\begin{equation}
\frac{\partial \phi(x,t)}{\partial t} = \int \mathcal{A}(\phi,x,y)\,dy + \frac{\eta(x,t)}{\sqrt{K}},
\label{SDE_general}
\end{equation}
where $\eta(x,t)$ is a Gaussian white noise with zero mean and with correlator
\begin{equation}
\langle \eta(x,t) \eta(x',t') \rangle = \int \mathcal{B}(\phi,x,y)\,dy \delta(x-x')\,\delta(t-t'),
\label{corr_gen}
\end{equation}
understood in the sense of It\=o. 

As described in Sec.~\ref{sectiondet} of the main text, we can obtain some insight into the transition into a multi-modal distribution by going over to a finite phenotypic space; specifically we assume that $-\pi < x \leq \pi$. We begin by analysing the deterministic dynamics of the model, which is found by letting $K \to \infty$ in Eq.~(\ref{SDE_specific}). That is,
\begin{eqnarray}
\label{detfield}
& & \frac{\partial \phi(x,t)}{\partial t} = -\int\phi(x,t)\phi(y,t)\frac{g(x-y)}{f(x)}\,dy  \\
&+& \int\frac{\phi(y,t)\phi(z,t)}{\phi * m(y)}m(y-z)r(x-(y+z)/2)\,dy\,dz.\nonumber
\end{eqnarray}
We substitute 
\begin{equation}
\label{LSA}
\phi(x,t) = 1 + \tilde{\phi}(x,t),
\end{equation}
into Eq.~(\ref{detfield}) and only keep linear terms in $\tilde{\phi}(x,t)$. Assuming that $g, m$ and $r$ are periodic with period $2\pi$ and normalised to unity in the interval $(-\pi, \pi)$, and $f(x)=1$, one finds that
\begin{equation}
\frac{d }{dt}\tilde{\phi}_k(t) = \left[ -1 - g_k + 2 r_k m_{-k/2} - r_k m_k m_{-k/2} \right] \tilde{\phi}_k(t).
\label{LSA_result}
\end{equation}
Here we have gone over to Fourier space, since the linear nature of the problem, and the translational invariance, make this a natural choice. The Fourier modes are defined by
\begin{equation}
h_k = \int^{\pi}_{-\pi} h(x) e^{-ikx}\,dx, \ \ \ \ 
h(x) = \frac{1}{2\pi}\,\sum_k h_k e^{ikx}.
\label{Fourier_def}
\end{equation}
From Eq.~(\ref{LSA_result}) we see that if the condition given in Eq.~(\ref{linearstab}) of the main text holds, then the homogeneous solution $\phi = 1$ is unstable.

If we wish to carry out a system-size expansion, as discussed in Sec.~\ref{weak}, then we write $\phi(x,t)=1+K^{-1/2}\xi(x,t)$ and expand in terms of $K^{-1/2}$. As discussed in the main text, this leads to Eq.~(\ref{LNA_result}):
\begin{equation}
\frac{d }{dt} \xi_k(t) = - \rho_k\xi_k(t) + \eta_k(t),
\label{LNA_SDE}
\end{equation}
where $\rho_k \equiv 1 + g_k - 2 r_k m_{-k/2} + r_k m_k m_{-k/2}$. Multiplying by
$e^{\rho_k t}$, this can be integrated to yield
\begin{equation}
\xi_k(t) = \xi_k(0)e^{-\rho_k t} + e^{-\rho_k t} \int^t_0 dt'\,e^{\rho_k t'} \eta_k(t').
\label{soln_SDE}
\end{equation}
Assuming that we begin with zero noise, $\xi_k(0)=0$, Eq.~(\ref{soln_SDE}) implies that
\begin{eqnarray}
& & \langle \xi_k (t) \xi_{-k}(t) \rangle = \exp{-(\rho_k + \rho_{-k})t}\,\times
\nonumber \\
& & \int^t_0 \int^t_0 dt' dt''\,\exp{(\rho_k t' + \rho_{-k} t'')} \langle \eta_k(t') \eta_{-k}(t'') \rangle. \nonumber \\
\label{corr_xi}
\end{eqnarray}
Using Eq.~(\ref{corr_k}) and letting $t \to \infty$, to obtain the result in the stationary state, one finds that if $(\rho_k + \rho_{-k}) > 0$,
\begin{equation}
\lim_{t \to \infty} \langle \xi_k (t) \xi_{-k}(t) \rangle = \frac{2 (2\pi)}{(\rho_k + \rho_{-k})}.
\label{stationary_state_result}
\end{equation}
\section{Numerical simulation of the individual-based model}
\label{appSim}
The numerical simulations of the individual-based model are performed using the Gillespie algorithm \cite{Gillespie77}. Before providing the details of our algorithm, we recall some basic elements of the process.  The state of the system at time $t$ is described by $N(t)$ real numbers, $x_i, i=1,\ldots,N(t)$, corresponding to the phenotypes of the $N(t)$ individuals present. The probability that individual $i$ initiates a reproduction event, given that a birth takes place, is $1/N(t)$ (since we consider no Allee effect, the reproduction probability is independent of the density of suitable mates); the probability that individual $i$ chooses individual $j$ to mate, given that individual $i$ is reproducing, is proportional to $m(x_i-x_j)$ (see Sec.~\ref{sec:Model}); finally, the probability density that the newborn individual has phenotype $x$, given that individuals $i$ and $j$ are reproducing, is given by $r(x-(x_i+x_j)/2)$ (again, see Sec.~\ref{sec:Model}).

There are two possible events (assuming $N(t)\geq 1$):
\begin{itemize}
\item[(i)] Death of an individual, with a rate (probability per unit of time) equal to $r_1=\int \gamma(x,\phi)dx=\sum_{i,j=1}^{N(t)} g(x_i-x_j)/Kf(x_i)$ (see Eqs.~(\ref{phi_density}) and (\ref{gamma_def})).
\item[(ii)] Birth of a new individual, with a rate $r_2=\int\beta(x,\phi)dx=\sum_{i,j=1}^{N(t)} m(x_i-x_j)/\sum_{l}m(x_i-x_l)\newline (=N(t))$ (see Eq.~(\ref{birthrate})).
\label{rates}
\end{itemize}
Therefore the probability that individual with phenotype $x_i$ (that we will denote as individual $i$) dies, given that a death event takes place, is proportional to $\sum^{N(t)}_{j=1} g(x_i-x_j)/Kf(x_i)$. 

The numerical simulations are, then, based on the following algorithm:
\begin{itemize}
 \item[1.] Set the initial state of the system, that is, the initial number of individuals, $N(t_0)$, their corresponding phenotypes, $x_i, i=1,\ldots,N(t)$, and the initial time, $t=t_0$.
\item[2.] Compute $r_1$ and $r_2$, as described earlier. Compute the time increment after which the next event takes place ($\Delta t$), which is an exponential random variable with average $(r_1+r_2)^{-1}$, so it can be computed as $\Delta t=-\ln(u)/(r_1+r_2)$, with $u$ a pseudo-random number with a uniform distribution in the interval $(0,1)$. Update the time $t=t+\Delta t$.
\item [3.] Establish what type of event takes place. With probability $r_1/(r_1+r_2)$ a death event takes place; go to 4a. Otherwise a birth event takes place; go to 4b.
\item [4a.] Establish which individual dies. Choose individual $i$ at random, with probability proportional to $\sum_{j=1}^{N} g(x_i-x_j)/Kf(x_i)$. Eliminate individual $i$. Update $N, N=N-1$. If $N=0$ the population becomes extinct and the simulation ends.
\item [4b.] Set the phenotype of the new individual. Choose individual $i$ uniformly at random to initiate reproduction. Choose individual $j$ to mate, at random with probability proportional to $m(x_i-x_j)$. Set the phenotype of the new individual as $x=(x_i+x_j)/2+\zeta$, where $\zeta$ is a random variable with probability density function given by $r(x)$. Update $N, N=N+1$ 
\item [5.] Go to 2 or finish.
\end{itemize} 
When considering periodic boundary conditions, this has to be taken into account when computing the competition and mating functions as well as the phenotype of the new individual.

\end{appendix}


\end{document}